\documentclass{osa-article}
\journal{oe}

\begin{document}

\title{Self-interaction of ultrashort pulses in an epsilon-near-zero nonlinear material at the telecom wavelength}

\author{Jiaye Wu,\authormark{1} Boris A. Malomed,\authormark{2} H.Y. Fu,\authormark{3} and Qian Li\authormark{1,*}}

\address{\authormark{1}School of Electronic and Computer Engineering, Peking University, Shenzhen 518055, China\\
\authormark{2}Department of Physical Electronics, School of Electrical Engineering, Faculty of Engineering, and Center for Light-Matter Interaction, Tel Aviv University, Tel Aviv 69978, Israel\\
\authormark{3}Tsinghua-Berkeley Shenzhen Institute (TBSI), Tsinghua University, Shenzhen 518055, China}

\email{\authormark{*}liqian@ece.pku.edu.cn} 



\begin{abstract}
Dynamics of femtosecond pulses with the telecom carrier wavelength is investigated numerically in a subwavelength layer of an indium tin oxide (ITO) epsilon-near-zero (ENZ) material with high dispersion and high nonlinearity. Due to the subwavelength thickness of the ITO ENZ material, and the fact that the pulse's propagation time is shorter than its temporal width, multiple reflections give rise to self-interaction in both spectral and temporal domains, especially at wavelengths longer than the ENZ point, at which the reflections are significantly stronger. A larger absolute value of the pulse's chirp strongly affects the self-interaction by redistributing energy between wavelengths, while the sign of the chirp affects the interaction in the temporal domain. It is also found that, when two identical pulses are launched simultaneously from both ends, a subwavelength counterpart of a standing-wave state can be established. It shows robust energy localization in the middle of the sample, in terms of both the spectral and temporal intensity distributions.
\end{abstract}

\section{Introduction}
In the recent decade, the epsilon-near-zero (ENZ) materials have drawn much interest in studies of plasmonic metamaterials \cite{Naik2013} and photonics \cite{Liberal2017, Niu2018}. By tuning its permittivity to a near-zero value, the ENZ material can feature a refractive index much lower than 1 and other extraordinary optical properties, such as electromagnetic energy tunneling \cite{Silveirinha2006}, directive emission with invariable phase \cite{Enoch2002}, amplification of electric field  \cite{Campione2013}, enhancement of nonlinearity \cite{Ciattoni2016}, pulse shaping and tailoring \cite{Alu2007, Zhai2013}, slow-light trapping \cite{Ciattoni2013}, and the creation of confined ENZ modes \cite{Vassant2012, Campione2015}.

State-of-the-art material technologies for transparent conducting oxides (TCOs) \cite{Ray1983, Agura2003, Park2006, Naik2011} are widely used as ENZ materials, due to their wide ENZ spectral region, extending from ultraviolet to mid-infrared \cite{Naik2013, Niu2018}. The ENZ point, at which the real part of the permittivity falls to zero, can be tuned by modulating the dopant concentration. Based on TCO ENZ materials, ultrahigh nonlinear Kerr response \cite{Alam2016, Caspani2016}, all-optical switching \cite{Guo2017, Kinsey2015}, second-harmonic generation (SHG)  \cite{Capretti2015}, third-harmonic generation (THG) \cite{Luk2015}, and high-harmonic generation \cite{Yang2019} were all demonstrated. Among various sorts of TCOs, indium tin oxide (ITO), a common material used in displays, is often employed in studies of ENZ, due to its low cost, high nonlinearity \cite{Alam2016}, and near-infrared ENZ region \cite{Niu2018}.

Linear and nonlinear interactions of ultrashort optical pulses with matter were studied, severally, in near-subwavelength \cite{Zhai2013, Campione2013} and longer-than-wavelength \cite{Ciattoni2016} setups based on ideal ENZ materials. New results may be expected in a realistic setting which includes the subwavelength scale of the medium, near-zero permittivity, dispersion, Kerr nonlinearity, chirp, reflections and absorption.

In this work, the self-interactions of chirp-free and chirped ultrashort pulses in a highly dispersive and highly nonlinear ITO ENZ material are investigated numerically. The calculations are performed for 20 fs pulses at the telecom wavelength of 1.55 $\mu$m. Since the thickness of the ENZ material considered here is much smaller than the wavelength, the propagation time of the pulse is actually shorter than its temporal width, and multiple reflections from the sample's edges may result in self-interference and nonlinear self-interactions in the spectral and temporal regimes alike, especially for wavelengths longer than the ENZ wavelength. The interplay between the chirp, dispersion, and Kerr nonlinearity plays a major role in the observed dynamics. Larger absolute values of the chirp have a more significant impact on the spectral and temporal shapes, while the sign of the chirp affects only the temporal features.

A new standing-wave-like state can be established in the present setting, in terms of both spectral and temporal intensity distributions, when two identical ultrashort pulses are launched simultaneously from two edges of the sample. It is necessary to stress that the formation of standing waves via the classical interferences is not a relevant mechanism on the subwavelength scale, therefore we name patterns, produced by the interaction of counterpropagating pulses in the framework of the full system of Maxwell's equations, {\it quasi-standing-wave} (QSW) states. Potential applications, such as nanophotonic Fabry-P\'{e}rot ENZ resonators, can be designed by combining QSW patterns and scattering properties of ultrashort pulses on the subwavelength scale.

Thus, results reported in this work should help to understand the self- and cross-interaction of ultrashort pulses in subwavelength samples made of isotropic ENZ materials. The paper is structured as follows. In Section \ref{material}, the configuration of the ENZ material and its linear and nonlinear optical properties are introduced. In Section \ref{spulse}, the theoretical model and methods used for the modeling of ultrashort pulses are presented. Results for the self-interference and self-interactions of chirp-free and chirped pulses are reported in Sections \ref{nchirp} and \ref{ychirp}, respectively. In Section \ref{local}, the QSW state is demonstrated. The paper is concluded by Section \ref{conclude}.

\section{Model and theory \label{MaT}}
\subsection{ENZ material \label{material}}

The ITO material operating in the ENZ regime can be fabricated using magnetron sputtering and chemical vapor deposition. The value of the wavelength at the ENZ point may be controlled by selecting proper concentrations of the material constituents. The permittivity of ITO is accurately modelled by the Drude's formula for the complex relative permittivity, with real and imaginary parts $\varepsilon_{\rm r}$ and $\varepsilon_{\rm i}$ \cite{Drude1900}:

\begin{equation}
\displaystyle {\varepsilon}\left( \omega  \right) = \displaystyle {\varepsilon _{\rm{r}}} + i{\varepsilon _{\rm{i}}} = \displaystyle {\varepsilon _{\rm{b}}} - \frac{{\omega _{\rm{p}}^2}}{{\left( {{\omega ^2} + {\Gamma ^2}} \right)}} + i\frac{{\omega _{\rm{p}}^2\Gamma }}{{\left( {{\omega ^2} + {\Gamma ^2}} \right)\omega }}.
\end{equation}

\begin{figure}[htbp]
\centering
{\includegraphics[width=0.67\linewidth]{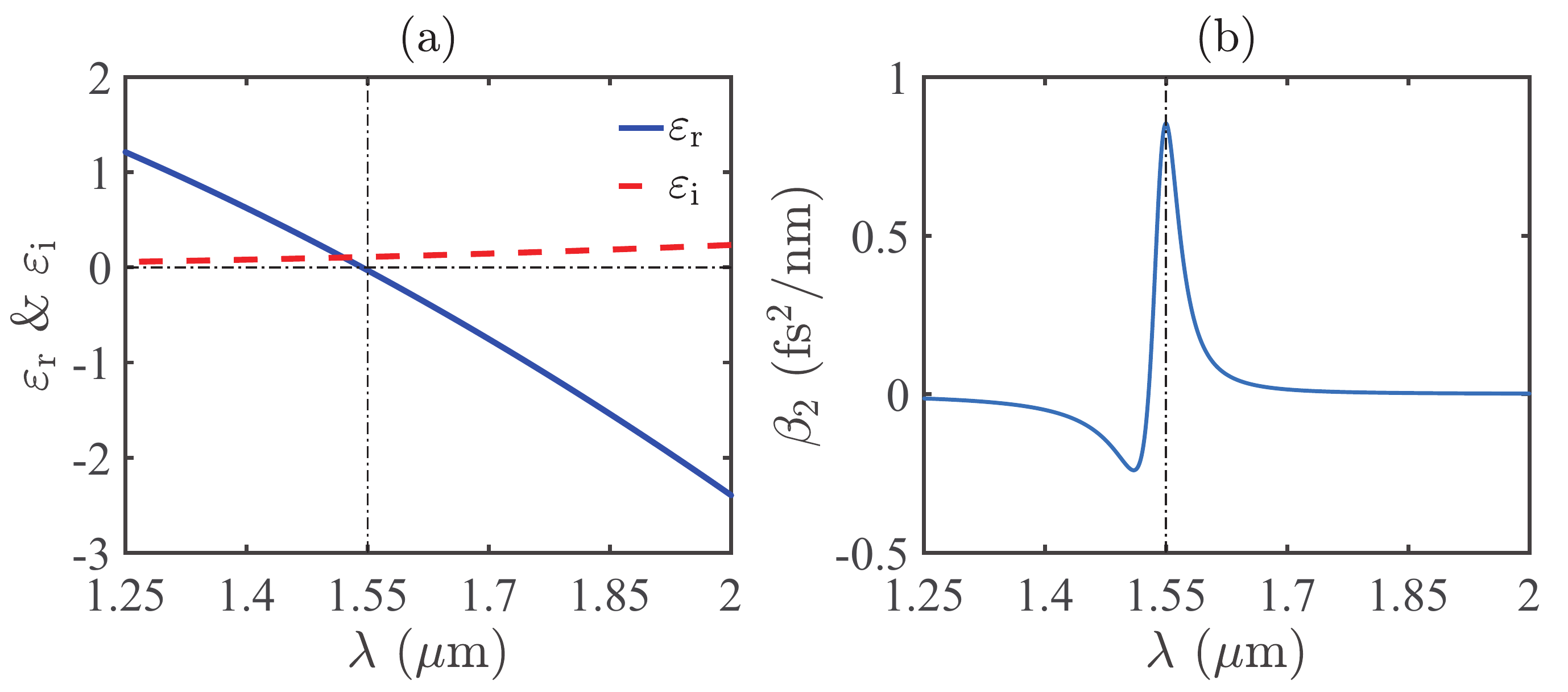}}
\caption{(a) The variation of the real and imaginary parts (the solid and dashed lines, respectively) of the complex permittivity, as a function of the wavelength. (b) The group-velocity dispersion, $\beta_2$, near $\lambda_{\rm C} = 1.55$ $\mu$m.}
\label{f1}
\end{figure}

Here $\varepsilon_{\rm b} = 3.528$ is the high-frequency (background) permittivity \cite{Naik2013, Ray1983}, and the Drude damping rate is $\Gamma$ = 0.1550 eV \cite{Naik2013, Ray1983}. High group velocity dispersion (GVD) at the telecom wavelength may be maintained by the ITO carrier density at the level of $6.30\times10^{20}$ cm$^{-3}$, which corresponds to plasma frequency $\omega_{\rm p}$ is $1.22\times10^{15}$ rad/s. These values ensure that the maximum GVD value, $\beta_2 = 0.85$ fs$^2$/nm, is attained at the central wavelength of $\lambda_{\rm C} = 1.55$ $\mu$m, and the resulting cross-over (ENZ) point, where $\varepsilon_{\rm r}$ vanishes, is $\lambda_{\rm ENZ} = 1.543$ $\mu$m. The variations of $\varepsilon_{\rm r}$ and $\varepsilon_{\rm i}$ near $\lambda_{\rm C} = 1.55$ $\mu$m are shown in Fig. \ref{f1}(a), and $\beta_2$ is plotted in Fig. \ref{f1}(b).

\begin{figure}[htbp]
\centering
{\includegraphics[width=0.6\linewidth]{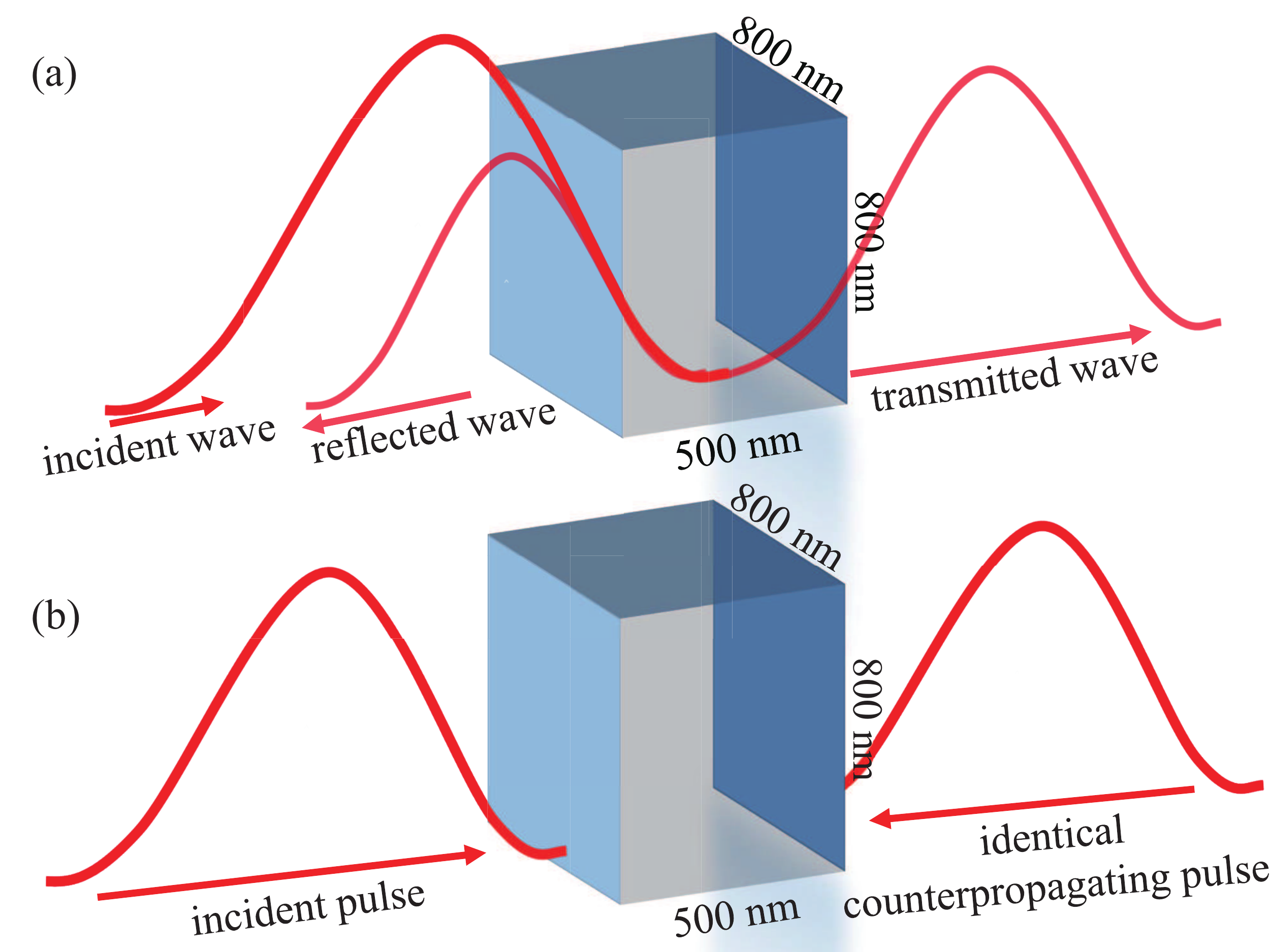}}
\caption{Schematic diagram of (a) a single ultrashort pulse incident onto the ITO ENZ material at the telecom wavelength; (b) the setup of counterpropagating pulses used to establish the QSW ({\it quasi-standing-wave}) state.}
\label{f2}
\end{figure}

The ITO sample is shaped as a subwavelength slab, as shown in Fig. \ref{f2}. The incident wave with $\lambda_{\rm C} = 1.55$ $\mu$m and the corresponding reflected and transmitted waves are schematically displayed in Fig. \ref{f2}(a), and the setup to establish the QSW is presented in Fig. \ref{f2}(b). To maintain the subwavelength character of the expected self-interactions and their adequate numerical simulations, a sufficient propagation length is necessary. Therefore, the thickness of the slab is taken to be 500 nm, which is around 1/3 of the central wavelength of the pulse. The known Kerr nonlinear index of ITO can be as high as $n_2 = 2.58\times10^{-16}$ m$^2$/W \cite{Alam2016}. Along with large $n_2$, a small cross-section area is necessary to secure the presence of the strong nonlinear effect. Accordingly, the cross section is fixed to be 800 nm $\times$ 800 nm, which also allows the bulk of the pulse energy to be transmitted in the propagation mode. With this cross-section, the corresponding self-phase modulation coefficient $\gamma$ is $1.63\times10^3$ W$^{-1}$m$^{-1}$, which is several orders of magnitude larger than in highly-nonlinear fibers, cf. \cite{Hansryd2001, Parmigiani2006, Salim2018, Lopez-Ripa2019}.

Scattering properties of the material in the linear regime are modelled by the transfer-matrix method. For normal incidence, the form of the transfer matrix is defined by relation

\begin{equation}
\left[ \begin{gathered}
  B \hfill \\
  C \hfill \\ 
\end{gathered}  \right] = \left[ {\begin{array}{*{20}{c}}
  {\cos \delta }&{{i\sin \delta }}/{n} \\ 
  {in\sin \delta }&{\cos \delta } 
\end{array}} \right]\left[ \begin{gathered}
  1 \hfill \\
  {n_s} \hfill \\ 
\end{gathered}  \right],
\label{e2}
\end{equation}

\noindent where $B$ and $C$ are parameters that characterize the layer's transmission, reflection, and absorption in Eq. (\ref{e4}) (see below), and $\delta = 2\pi nd /\lambda$ is the phase thickness, related to the physical thickness d and the wavelength. Further, $n_s$ is the refractive index of the substrate (in the present context, it is also the refractive index of the ambient medium, which may be air or vacuum), while $n = n_{\rm r} +ik$ is the complex refractive index characterized by real part $n_{\rm r}$ and extinction coefficient $k$, which are related to complex permittivity $\varepsilon$:

\begin{equation}
\begin{gathered}
  {n_{\rm{r}}} = \sqrt {\frac{{\sqrt {{\varepsilon _{\rm{r}}}^2 + \varepsilon _{\rm{i}}^2}  + {\varepsilon _{\rm{r}}}}}{2}} , \hfill \\
  k = \frac{{{\varepsilon _{\rm{i}}}}}{{2{n_{\rm{r}}}}}{\text{ = }}\sqrt {{n_{\rm{r}}}^2 - {\varepsilon _{\rm{r}}}} . \hfill \\ 
\end{gathered}
\label{e3}
\end{equation}

\begin{figure}[htbp]
\centering
{\includegraphics[width=0.5\linewidth]{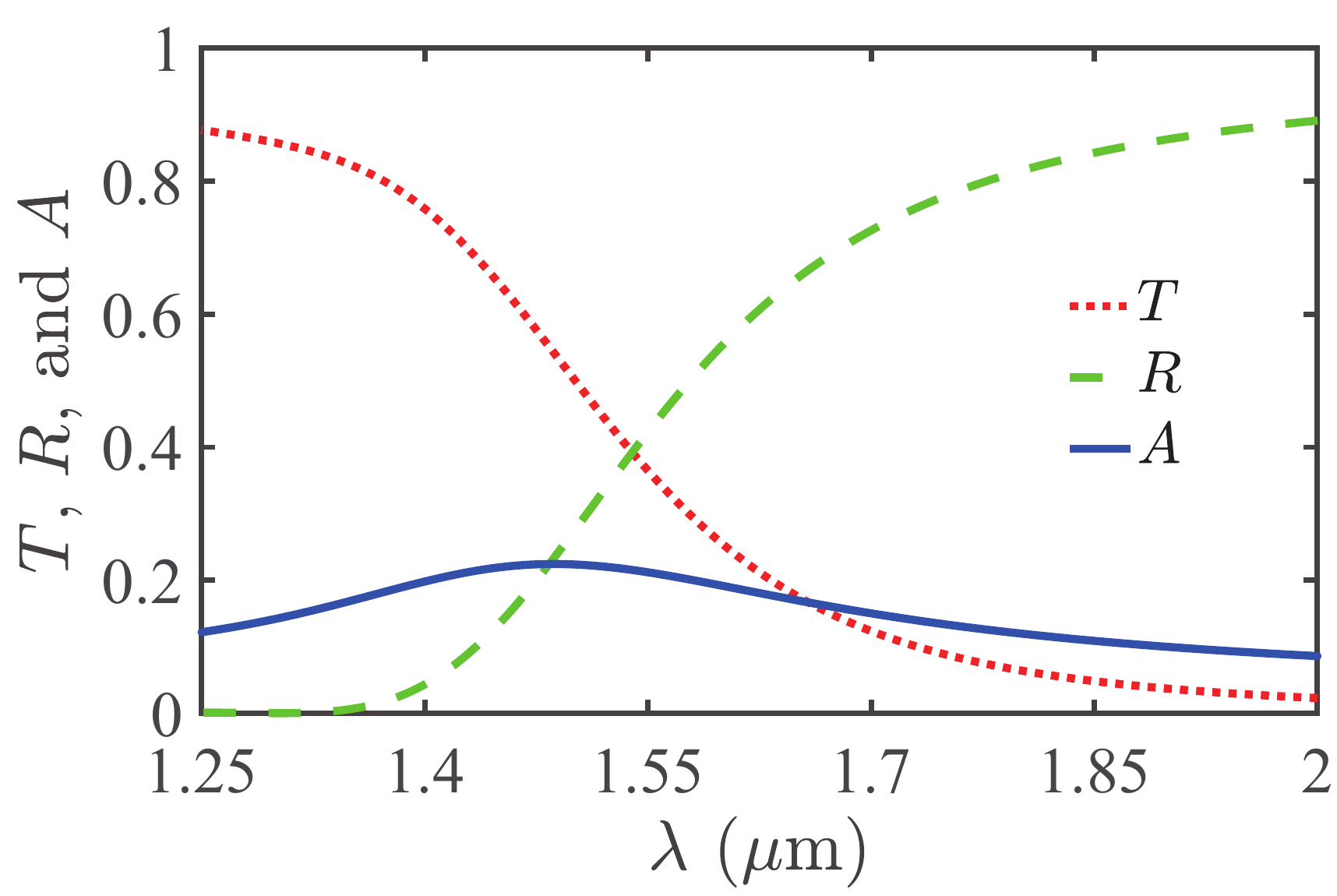}}
\caption{The transmission (dashed-dotted line), reflection (dashed line), and absorption (solid line) coefficients of the ITO sample under the consideration.}
\label{f3}
\end{figure}

The reflection, transmission, and absorption coefficients are determined by the transfer matrix in Eq. (\ref{e2}) as follows:

\begin{equation}
\begin{gathered}
  R = \left( {\frac{{{n_s}B - C}}{{{n_s}B + C}}} \right){\left( {\frac{{{n_s}B - C}}{{{n_s}B + C}}} \right)^*}, \hfill \\
  T = \left( {1 - R} \right)\psi , \hfill \\
  A = \left( {1 - R} \right)\left( {1 - \psi } \right), \hfill \\ 
\end{gathered} 
\label{e4}
\end{equation}

\noindent where $\psi = n_s / {\rm Re}(BC^*)$ is the potential transmittance, Re is for the real part of complex expression, and $*$ stands for the complex conjugate. The results obtained for the 500-nm ITO sample are shown in Fig. \ref{f3}, where the transmission, reflection, and absorption coefficients are depicted by the dotted, dashed, and solid lines, respectively. At wavelengths shorter than the values at the ENZ point, ITO exhibits high transmission and low reflection, while at longer wavelengths the opposite situation occurs. The form of these curves agrees with what is expected for an ideal ENZ material \cite{Zhai2013}.

\subsection{Modeling the ultrashort pulses \label{spulse}}

In this work, an ultrashort input pulse is taken as

\begin{equation}
u = \sqrt {{P_0}} \cos \left( {{\omega _0}t - \frac{{C{t^2}}}{{2T_0^2}}} \right)\exp \left( { - \frac{{{t^2}}}{{2T_0^2}}} \right),
\label{e5}
\end{equation}

\noindent where $t$ is time, $\omega_0$ the central frequency, $C$ the chirp, and $T_0$ the temporal width. The FWHM temporal size is taken as $T_{\rm FWHM} = 2(\ln2)^{1/2}T_0 = 20$ fs, and the peak power is chosen to be $P_0 = |\beta_2|/(T_0^2\gamma) = 1.31$ kW, which maintains the balance between the GVD and nonlinearity in the initial pulse. The evolution of the optical fields in the nonlinear nonmagnetic material is governed by the Maxwell's equations \cite{Agrawal2013}:

\begin{equation}
\begin{gathered}
  \nabla  \times {\mathbf{H}} = \frac{{\partial {\mathbf{D}}}}{{\partial t}}, \hfill \\
  {\mathbf{D}} = {\varepsilon _0}{\varepsilon _{\rm{R}}}{\mathbf{E}} + {{\mathbf{P}}_{\rm{L}}} + {{\mathbf{P}}_{{\rm{NL}}}}, \hfill \\
  \nabla  \times {\mathbf{E}} =  - \frac{{\partial {\mathbf{B}}}}{{\partial t}}, \hfill \\
  {\mathbf{B}} = {\mu _0}{\mathbf{H}}, \hfill \\ 
\end{gathered}
\label{e6}
\end{equation}

\noindent where $\bf E$ and $\bf H$ are electric and magnetic field vectors, $\bf D$ and $\bf B$ electric and magnetic inductions, $\varepsilon_0 = 8. 85\times10^{-12}$ F/m and $\mu_0 = 4\pi\times10^{-7}$ N/A$^2$ are the vacuum permittivity and permeability, while ${\bf P}_{\rm L}$ and ${\bf P}_{\rm NL}$ stand for the linear and nonlinear parts of the induced electric polarization in the material. ${\bf P}_{\rm L}$ is determined by the linear refractive index $n$, and experimental data for Kerr nonlinearity in ITO \cite{Alam2016} is used to define ${\bf P}_{\rm NL}$ as follows when only the third-order nonlinearity is considered. 

\begin{equation}
{{\bf{P}}_{{\rm{NL}}}} = \frac{{4c\varepsilon _0^2{n^2}}}{3}{n_2} \vdots {\bf{E}}\left( {{\bf{r}},t} \right){\bf{E}}\left( {{\bf{r}},t} \right){\bf{E}}\left( {{\bf{r}},t} \right),
\end{equation}

\noindent where $c$ is the speed of light in vacuum, $\varepsilon_0$ the vacuum permittivity, $n$ is the refractive index obtained by Eq. (\ref{e3}), $n_2$ the Kerr nonlinear index, and $\bf E$ the electric field. The Maxwell's equations were solved numerically by means of the finite-difference time-domain method.

\section{Self-interactions of chirp-free pulses \label{nchirp}}

\begin{figure}[tbp]
\centering
{\includegraphics[width=0.8\linewidth]{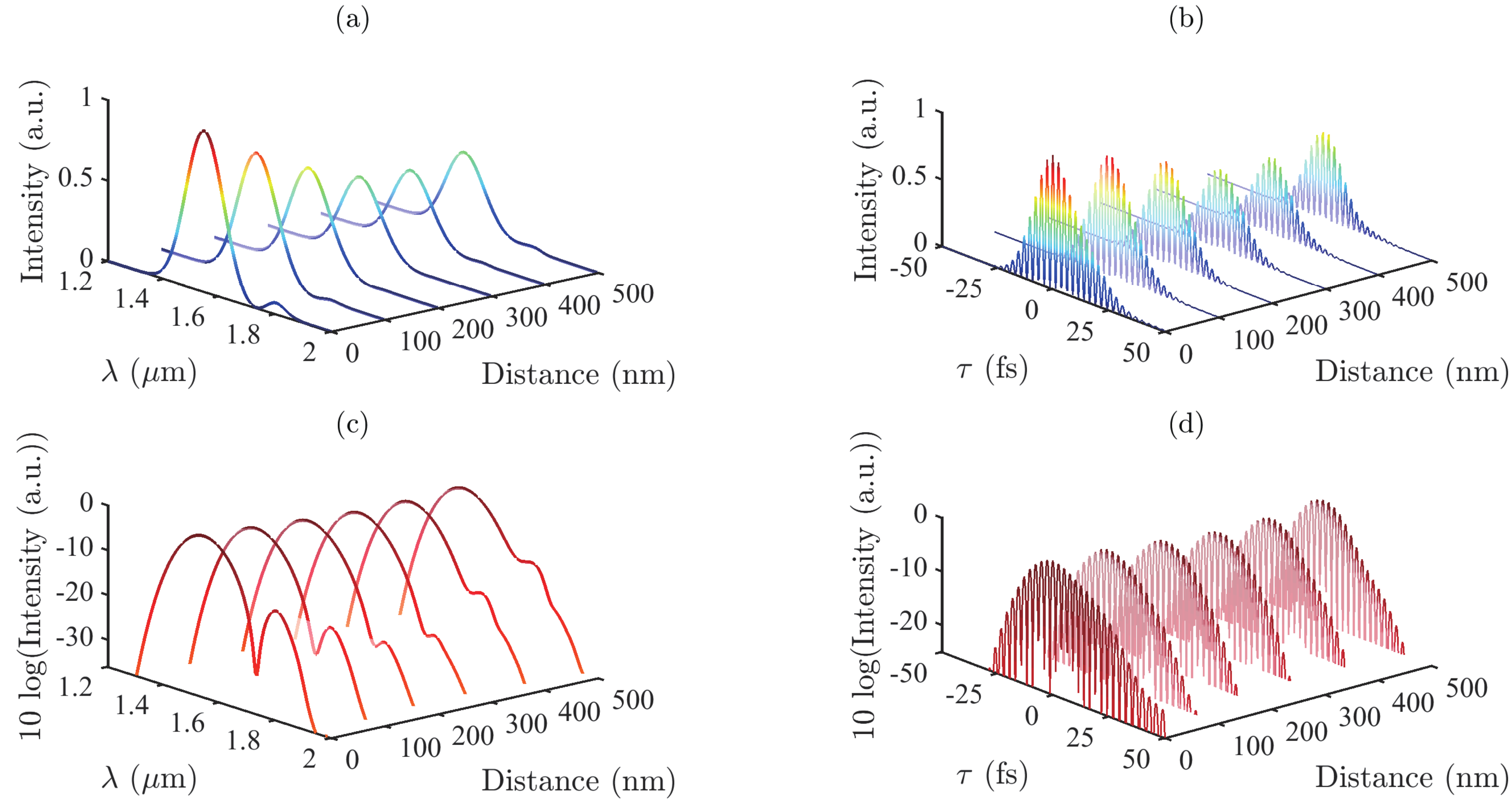}}
\caption{The evolution of the intensity of the electromagnetic field, produced by direct simulations of the Maxwell's equations Eq. (\ref{e6}) in the spectral (a) and temporal (b) domains. Panels (c) and (d) show the same pictures on the logarithmic scale.}
\label{f4}
\end{figure}

First, a chirp-free ($C = 0$) 20-fs Gaussian pulse with the central wavelength of 1.55 $\mu$m is launched into the sample. Several numerical monitors are placed along the incident direction, to record the fields and spectra in the corresponding cross-section, until all the signals leave the optical structure. The results obtained in the spectral and temporal domains are plotted in Figs. \ref{f4}(a) and \ref{f4}(b), respectively, and are additionally displayed on the logarithmic scale in Figs. \ref{f4}(c) and \ref{f4}(d). Labels ``0 nm'' to ``500 nm'' are distance marks within the sample, along the propagation direction, starting from the incident cross-section. Due to the fact that the pulse propagation in such a subwavelength structure is not unidirectional, Figs. \ref{f4}(a)-\ref{f4}(d) display combined results, produced by the superposition of waves arriving from different directions, taking into regard multiple reflections and the accumulated nonlinearity. The extremely short propagation time and large GVD-induced velocity difference among different wavelength components make it impossible to distinguish the initial wave and contributions produced by reflections. Therefore, Fig. \ref{f4} cannot be interpreted as a result of unidirectional pulse propagation. Nevertheless, it displays a relevant physical picture, showing the actual distribution of the optical fields in the sample. It is seen in Fig. \ref{f4} that, in general, the spectral and temporal intensity gradually decays due to the intrinsic absorption, determined by coefficient $A$, which is shown in Fig. \ref{f3}. At the output end of the sample (which corresponds to the 500 nm propagation length), the Fresnel effect on the material-air interface results in slight increase of the spectral intensity and the appearance of a temporal trailing edge.

As mentioned above, the results displayed in Fig. \ref{f4} are not true pulses, because the thickness of the sample is smaller than the central wavelength, and the propagation time is shorter than the temporal width of the pulse. Therefore, an essential contribution to the patterns in Fig. \ref{f4} is self-interaction in both spectral and temporal domains. As concerns the spectra, since the material is isotropic, the $T$, $R$, and $A$ coefficients produced by Eq. (\ref{e4}) pertain equally well to all propagation directions, which means the reflection may happen many times, until the entire pulse exits the structure. This situation resembles that found in studies of the Lorentz-Duffing film \cite{Brio2019}, and is especially relevant to wavelengths longer than one at the ENZ point. On the other hand, the thickness of the sample is merely 1/3 of the central wavelength, which indicates that the incident leading edge of the pulse is reflected back and forth before the whole pulse will pass through the structure. The reflected part interferes with the rest of the pulse linearly, and interacts with it nonlinearly, resulting in the secondary intensity peak at longer wavelengths, which is clearly seen in Figs. \ref{f4}(a) and \ref{f4}(c). The self-interaction manifests itself in the temporal domain as well. The calculations demonstrate that the slowest group velocity is 34.53 nm/fs, corresponding to $\lambda = 1.529$ $\mu$m. With this speed, the pulse's component needs 14.48 fs to travel through the structure, which is shorter than the pulse's temporal width, 20 fs. When multiple reflections are considered, the time needed for the entire energy to leave the sample is much longer than 20 fs. Therefore, although the designed ENZ sample has the subwavelength size, the actually available length and time are sufficient for the accumulation of nonlinear effects.

\section{Self-interactions of chirped pulses \label{ychirp}}

\begin{figure}[tbp]
\centering
{\includegraphics[width=\linewidth]{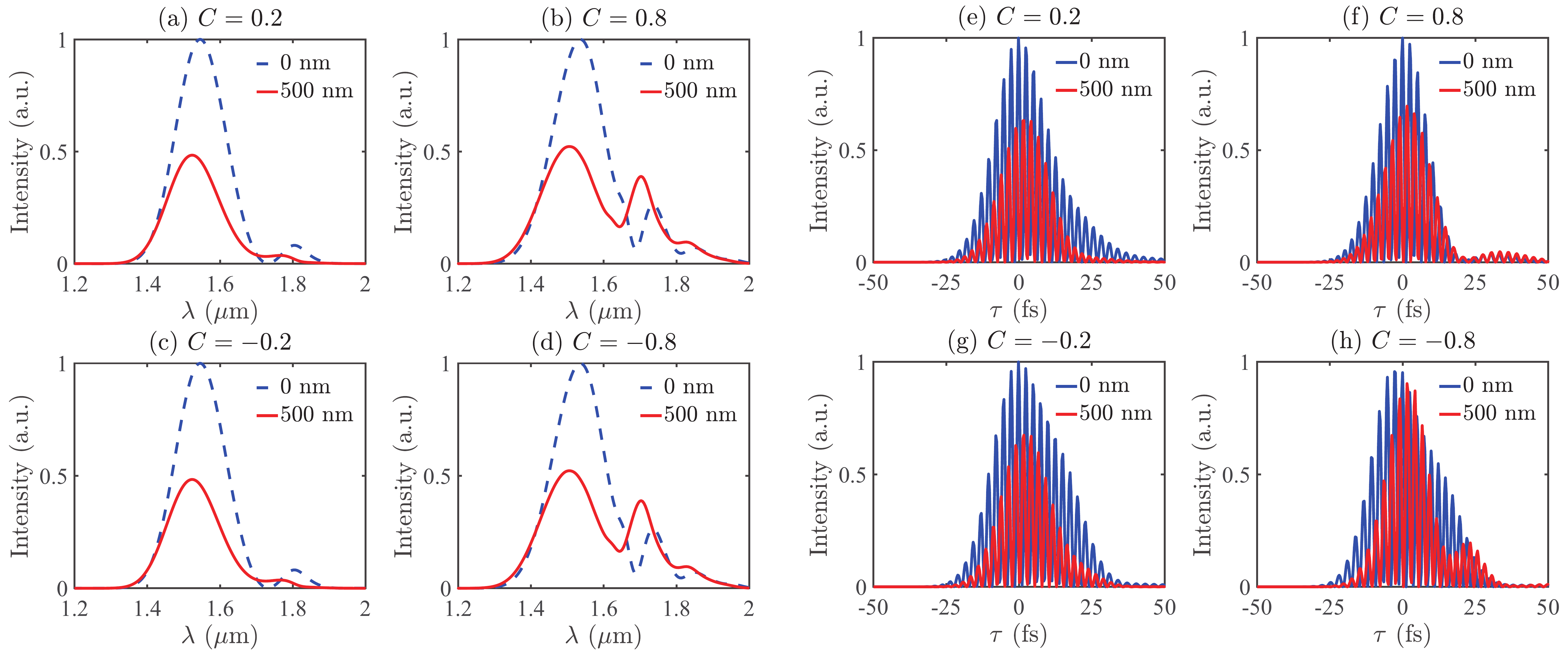}}
\caption{The spectra for the propagation distance 0 nm (the dashed line) and 500 nm (the solid line), with the initial chirp in Eq. (\ref{e5}) $C = 0.2$ (a), $0.8$ (b), $-0.2$ (c), $-0.8$ (d). (e)-(h) are the same as in (a)-(d), but for the intensity distribution in the temporal domain.}
\label{f5}
\end{figure}

It is well known that the interplay of the pulse's chirp with the GVD and Kerr nonlinearity plays a significant role in spectral and temporal shaping of the pulses \cite{Li2009, Wu2019}. We have considered different values of the chirp in input \ref{e5}, $-0.8 < C < +0.8$, to investigate their impact on the self-interaction. In Fig. \ref{f5}, we display the results for relatively small and large chirps, viz., $C = \pm0.2$ and $C = \pm0.8$, respectively.

It is seen in Figs. \ref{f5}(a)-\ref{f5}(d) that positive and negative chirps with equal absolute values generate identical spectra, a larger chirp having a more significant impact on longer-wavelength components, making the secondary spectral peak stronger. The secondary peak is a result of stronger reflections at long wavelengths, see Fig. \ref{f3}.

However, the sign of the chirp matters for the temporal intensity, as seen in Figs. \ref{f5}(e)-\ref{f5}(h), which shows that a negative chirp tends to shift the optical energy to the trailing (right) part in the temporal pattern. This trend is explained by the fact that a negative chirp in Eq. (\ref{e5}) causes the signal to oscillate at a higher frequency in the trailing edge of the pulse, which indicates a higher energy density.

\begin{figure}[htbp]
\centering
{\includegraphics[width=0.5\linewidth]{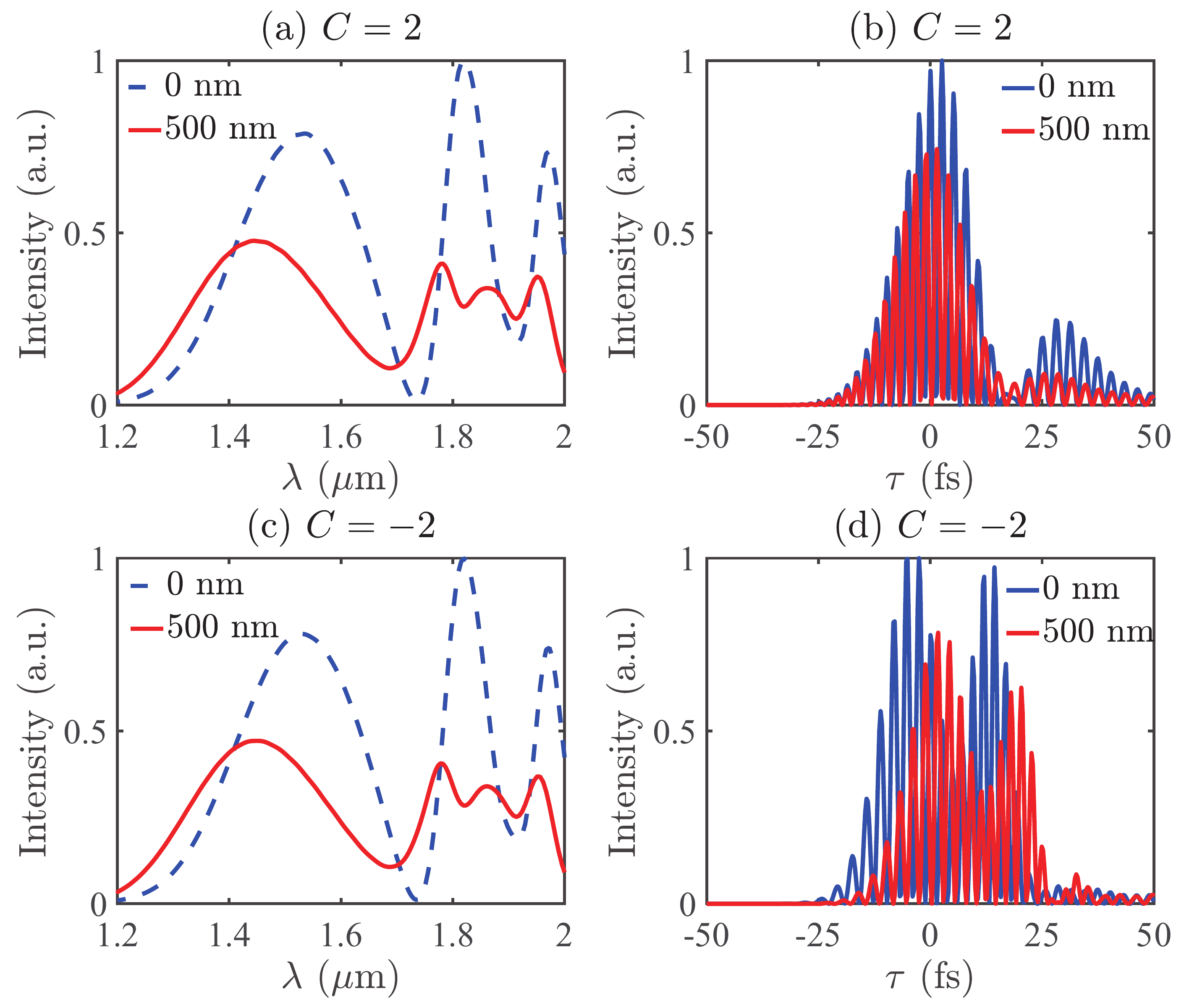}}
\caption{The spectral intensity distribution with chirps (a) $C = 2$ and (c) $C = -2$ for the propagation distance 0 nm (the dashed line) and 500 nm (the solid line). The temporal intensity distribution with chirps (b) $C = 2$ and (d) $C = -2$.}
\label{f6}
\end{figure}

A natural trend, demonstrated by Figs. \ref{f5}(a)-\ref{f5}(h), is that the larger chirp causes a stronger deformation of the optical patterns. Additional simulations demonstrate that the chirp with a still larger absolute value leads to deeper splitting of the spectral and temporal patterns into two peaks and, eventually, into several peaks. This can be seen in the spectral and temporal results of Fig. \ref{f6}, in which case we set $C = \pm2$. It is seen in Figs. \ref{f6}(a) and \ref{f6}(c) that, the for a very large chirp, the spectra have already split into three peaks (at the input, corresponding to propagation distance 0 nm) and four peaks ( for distance 500 nm). The temporal splitting is more significant with a negative chirp than a positive one, as can be interpreted from Figs. \ref{f6}(b) and \ref{f6}(d). This trend will become more and more significant when the chirp becomes larger, and eventually either the spectral or the temporal shape (or both) will deform completely.

\section{QSW (quasi-standing-wave) patterns \label{local}}

\begin{figure}[htbp]
\centering
{\includegraphics[width=0.8\linewidth]{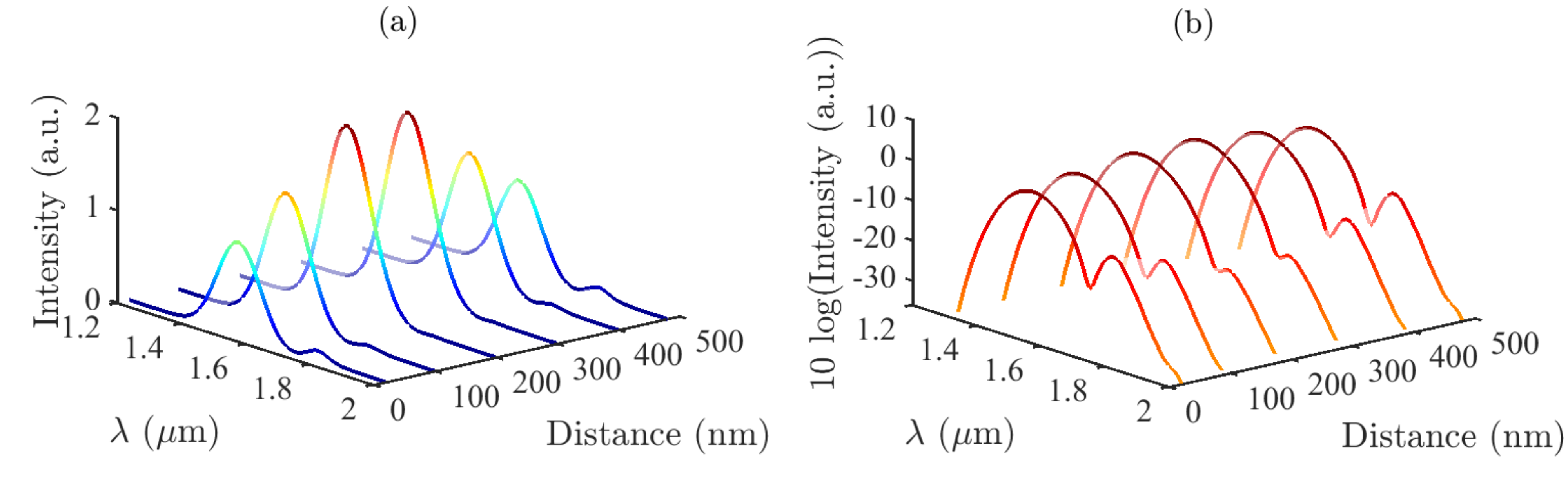}}
\caption{(a) The spectral intensity distribution corresponding to the QSW (quasi-standing-wave) pattern, created by the interplay of two pulses launched in opposite directions. (b) The same on the logarithmic scale.}
\label{f7}
\end{figure}

The designed subwavelength sample, made of the ITO ENZ material, can also support standing-wave-like states, featuring energy localization in both spectral and temporal realms. To achieve this state, we ran the simulations with two identical 20-fs chirp-free pulses launched into the sample from opposite edges, as shown in Fig. \ref{f2}(b). The resulting spectra are shown in Fig. \ref{f7}.

\begin{figure}[htbp]
\centering
{\includegraphics[width=0.8\linewidth]{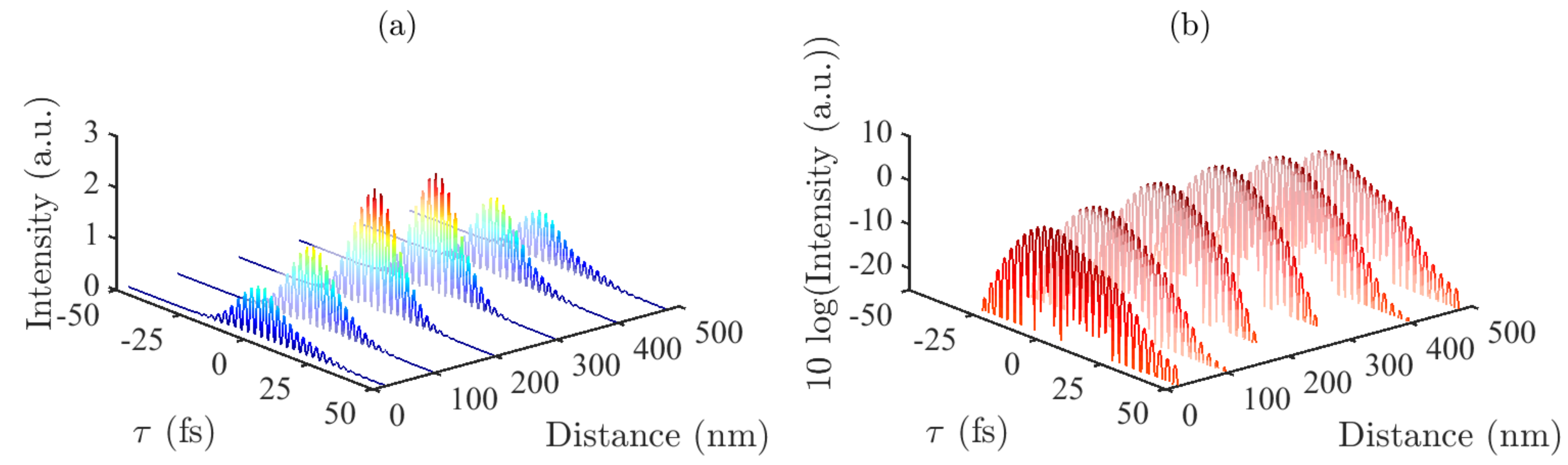}}
\caption{Panels (a) and (b) display the temporal-domain counterparts of the spectral image of the QSW pattern shown in Fig. \ref{f7}.}
\label{f8}
\end{figure}

It is seen that the spectral energy density, corresponding to the standing-wave pattern, features a stable peak precisely in the middle of the sample. In comparison to Fig. \ref{f4}, the spectrum observed in Fig. \ref{f7} {\it does not} amount to straightforward superposition of its counterparts produced, separately, by two individual input pulses. However, as mentioned above, this dynamical pattern, built on the subwavelength scale, cannot be a ``genuine'' standing wave, as the classical interference does not take place on such scales. Instead, it is a more complex structure, produced by the interplay of two counterpropagating subwavelength inputs, under the action of all physical effects included in the underlying system of the Maxwell's equations.

On the other hand, it is worthy to note that, as seen in Fig. \ref{f7}(b), the nonlinearity and ENZ reflections induce a secondary long-wave peak, similar to what was observed above in the case of the single-pulse input in Fig. \ref{f5}(d).

The temporal-domain image of the quasi-standing-wave pattern is displayed in Fig. \ref{f8}, which also demonstrates the localization of energy exactly in the middle of the sample. Combining the high reflection at wavelengths above the ENZ point and the localization features observed in Figs. \ref{f7} and \ref{f8}, a metamaterial-based Fabry-P\'{e}rot resonator may be designed, operating not on the classical interference, but rather on the interaction of subwavelength pulses.

It is also worthy to mention that the second- and third-order nonlinear susceptibilities of ITO near the ENZ point make it possible to implement SHG and THG effects in this material \cite{Capretti2015, Luk2015}. In this connection, the present results suggest that the energy of longer-wavelength incident light can be stored and accumulated in the sample by means of the energy-localization mechanism demonstrated here. Then, the accumulated energy may be used as a source for the generation of the second and third harmonics carried by the shorter wavelengths. Subsequent emission of these harmonics will be facilitated by the presence of the broad transmission window, as seen in Fig. \ref{f3}.

Additionally, regarding the robustness of the results presented in Sections \ref{nchirp}, \ref{ychirp}, and \ref{local}, for ultrashort pulses with similar temporal widths at tens of femtoseconds, e.g., at the level of 30 fs or 50 fs, the dynamics will be basically the same as shown here. However, when the pulse width is extended to a larger value, such as 200 fs (the width an order of magnitude longer), the dynamical patterns are totally different, due to the mismatch between the subwavelength and sub-temporal-width characteristics. Actually, the 200 fs pulse looks like a continuous wave, in comparison with the 20 fs one. When the pulse width is too large, although the subwavelength setup is maintained, the spectrum would become too narrow for nonlinear interactions, while the sub-temporal-width single-trip propagation time is completely irrelevant. Therefore, the underlying mechanisms and ensuing dynamics are expected to be very different.

\section{Conclusions \label{conclude}}

We have investigated the spectral and temporal self-interaction of chirp-free and chirped ultrashort pulses in a highly dispersive and highly nonlinear ITO ENZ material, by means of systematic simulations of the full system of the Maxwell's equations in the material medium at the telecom carrier wavelength of 1.55 $\mu$m. Due to the subwavelength thickness of the sample and propagation time which is shorter than the pulse's temporal width, multiple reflections drive complex interplay between the dispersion and Kerr nonlinearity, in the spectral and temporal domains alike. Effects of the chirp were considered too, with the conclusion that a larger absolute value of chirp has a more significant impact on the spectral and temporal shapes of the field, leading to the generation of a long-wavelength peak, and eventually splitting into two or larger number of well-pronounced peaks, while the sign of the chirp affects only the temporal shape.

Additionally, a subwavelength counterpart of the formation of standing waves was investigated, by simulating the interaction of two pulses launched into the sample from the opposite edges. The so established stable pattern demonstrates energy localization in the middle of the sample, in the spectral and temporal domains alike. These results may be used for the design of nanoscale counterparts of the Fabry-P\'{e}rot resonator, using the ENZ material.

Thus, this work suggests a new perspective for the studies of light-matter interactions, as well as self- and cross-interactions between ultrashort pulses, in nonlinear nanophotonic settings.

\section*{Funding}
National Natural Science Foundation of China (61675008); Shenzhen Science and Technology Innovation Commission (GJHZ20180411185015272, KQJSCX20170727163424873).

\section*{Acknowledgments}
J.W. is indebted to Dr. K. Nakkeeran for useful discussions.  B.A.M. appreciates hospitality of School of Electronic and Computer Engineering at the Peking University Graduate School in Shenzhen. 

\section*{Disclosures}
The authors declare no conflicts of interest.  

\bibliography{References}

\end{document}